\documentstyle[12pt]{article}
\raggedright 
\renewcommand{\section}[1]{\refstepcounter{section}
\vspace{24pt}\noindent{\bf\arabic{section}.\quad #1}
\vspace*{12pt}}
\newcommand{\ulsect}[1]{\vspace{18pt}\noindent{\bf #1}
\vspace*{12pt}}

\begin{document}

\begin{flushright} CERN-TH.7034/93\\
\end{flushright}
\vspace*{10mm}
\begin{center}
{\bf The high-frequency finite-temperature quark dispersion relation}\\[10mm]
David Seibert$^*$\\[5mm] Theory Division, CERN, CH-1211 Geneva 23,
Switzerland\\[10mm]
{\bf Abstract}\\
\end{center}
\hspace*{12pt}I calculate the dispersion relation for quarks of mass
$m$ and momentum $k$ in a quark gluon plasma at temperature $T$, in
the limit $m^2+k^2 \gg (gT)^2$, where $g$ is the strong coupling constant.
I find three contributions to the dispersion relation: one that depends
on $T$ but not $m$ or $k$, one that depends on $m$ and $T$ but not $k$,
and third contribution that depends on all three (and is opposite in sign
to the other two).\\
\vfill
\begin{center}
{\em Submitted to Phys.\ Rev.\ D}
\end{center}
\vfill
CERN-TH.7034/93\\
October 1993
\vspace*{10mm}
\footnoterule
\vspace*{3pt}
\noindent$^*$Address after October 12, 1993:
Physics Department, Kent State University, Kent, OH 44242 USA.
Internet: seibert@ksuvxa.kent.edu.\\
\newpage\setcounter{page}{1} \pagestyle{plain}
     \setlength{\parindent}{12pt}


Recently, there has been much interest in calculating the production rates
for both massless and massive quarks in ultrarelativistic nuclear
collisions [\ref{rmueller}-\ref{rcharm}].  Unfortunately, there have been
no rigorous thermal field theory calculations to compare to approximate
results, probably because the ``naive'' production rates diverge for
massless quarks.  The production rates can be regularized by summing over
hard thermal loops [\ref{rKW}, \ref{rW}], following the treatment of
Braaten and Pisarski [\ref{rBP}].  However, this is a difficult calculation,
and has not been done so far.

The closest approach to a full calculation has been to treat the thermal
quark masses for massless quarks as if they were bare masses, in order to
regulate the divergent production rates [\ref{rmueller}].  However, the
thermal corrections for massive quarks are not included, and these can be
important when $m \sim T$.  There have been several papers in which the
thermal dispersion relation was calculated for massive quarks
[\ref{rlb}-\ref{rbo}], but these have concentrated on the low-frequency
dispersion relation.  Also, no calculations have been done for systems out
of chemical equilibrium, while in the early stages of an ultrarelativistic
nuclear collision the quarks (and anti-quarks) have densities below their
chemical equilibrium values.

Because the main interest here is calculating effective masses to regulate
production rates, I calculate the dispersion relation in the high-frequency
limit.  This is because the divergences in the rates occur at small values
of $t$ or $u$, but at arbitrary values of $s$, so the quarks involved have
momenta of order $(T^2+mT)^{1/2}$ rather than of order $gT$.  I calculate
the thermal masses for arbitrary bare mass, following the calculation of
Petitgirard [\ref{rpg}] and working to order $g^2$, to provide a smooth
transition from the production of massless quarks to that of massive
quarks.


I begin with a quark gluon plasma of temperature $T$ with strong coupling
constant $g$.  To one-loop order, the quark self-energy $\Sigma(K)$ can
be written as
\begin{equation}
\Sigma(K) = -a\not\!\!{K} -b\not\!{u} -cm,
\end{equation}
where
\begin{eqnarray}
a &=& \frac {1} {4k^2}
\left[ \mbox{Tr} \left( \not\!\!{K} \mbox{Re} \Sigma \right)
- \omega \mbox{Tr} \left( \not\!{u} \mbox{Re} \Sigma \right) \right], \\
b &=& \frac {1} {4k^2}
\left[ (\omega^2-k^2) \mbox{Tr} \left( \not\!{u} \mbox{Re} \Sigma \right)
- \omega \mbox{Tr} \left( \not\!\!{K} \mbox{Re} \Sigma \right) \right], \\
c &=& \frac {1} {4m} \mbox{tr} \left( \mbox{Re} \Sigma \right).
\end{eqnarray}
Here the quark four-momentum $K = (\omega, \vec{k})$, $u$ is the matter
four-velocity and $m$ is the bare quark mass.  The Lorentz-invariant
denominator of the quark propagator is then
\begin{equation}
\mbox{Tr} \left( \not\!\!{K} -m -\Sigma \right)^2 = (1+a)^2K^2
+2(1+a)bK_{\mu}u^{\mu} +b^2 -(1-c)^2m^2,
\end{equation}
and its poles give the quark dispersion relation.

The dispersion relation in the plasma rest frame is thus
[\ref{rpg}],
\begin{equation}
(1+a) \omega + b = \sqrt{ (1+a)^2k^2 + (1+c)^2m^2 }. \label{ed0}
\end{equation}
Because $\Sigma = {\cal O} (g^2)$, I rewrite the dispersion relation
(\ref{ed0}) as
\begin{equation}
\omega = \sqrt{ k^2 + m^2 +2(c-a)m^2 - \frac {2b} {\omega} (k^2+m^2) }
+ {\cal O} (g^3), \label{ed1}
\end{equation}
for $\omega > k \gg gT$.  I then use the relation
\begin{equation}
\omega^2 = k^2 + m^2 + {\cal O} (g^2) \label{eid1}
\end{equation}
to obtain
\begin{equation}
\omega = \sqrt{ k^2 + m^2 +\delta } + {\cal O} (g^3), \label{ed2}
\end{equation}
where
\begin{equation}
\delta = 2(c-a)m^2 - 2b\omega = \frac m 2 \mbox{tr} \left( \mbox{Re} \Sigma
\right) +\frac 1 2 \mbox{Tr} \left( \not\!\!{K} \mbox{Re} \Sigma \right).
\end{equation}

The self-energy projections have been calculated by Petitgirard [\ref{rpg}],
\begin{eqnarray}
\mbox{Tr} \left( \not\!\!{K} \mbox{Re} \Sigma \right) &=& 2 g^2 C_F
\int_0^{\infty} \frac {p dp} {(2\pi)^2} \left[ \left( 4 + \frac {K^2+m^2}
{2pk} L_B(p) \right) n_B(p) \right. \nonumber \\
& & \left. + \left( 4p + \frac {K^2+m^2} {2k} L_F(p) \right)
\frac {n_F(E)} {E} \right], \\
\mbox{tr} \left( \mbox{Re} \Sigma \right) &=& \frac {4 g^2 C_F} {k}
\int_0^{\infty} \frac {p dp} {(2\pi)^2} \left[ -L_B(p) \frac {n_B(p)} {p}
+L_F(p) \frac {n_F(E)} {E} \right],
\end{eqnarray}
where $C_F$ is the quadratic Casimir invariant of the quark representation,
$n_{B(F)}$ is the Bose (Fermi) distribution function, $E=(p^2+m^2)^{1/2}$,
\begin{eqnarray}
L_B(p) &=& \ln \left| \frac {\left[ K^2-m^2 +2p(\omega+k) \right]
\left[ K^2-m^2 -2p(\omega-k) \right]} {\left[ K^2-m^2 +2p(\omega-k) \right]
\left[ K^2-m^2 -2p(\omega+k) \right]} \right|, \\[2pt]
L_F(p) &=& \ln \left| \frac {\left[ K^2+m^2 +2(E\omega+pk) \right]
\left[ K^2+m^2 -2(E\omega-pk) \right]} {\left[ K^2+m^2 +2(E\omega-pk) \right]
\left[ K^2+m^2 -2(E\omega+pk) \right]} \right|.
\end{eqnarray}
As the logarithms are already multiplied by $g^2$, and I am interested only
in the lowest order corrections to the dispersion relation, I take $K^2=m^2$
in the logarithms, which is equivalent to using the identity~(\ref{eid1}),
obtaining the simpler forms
\begin{eqnarray}
L_B(p) &=& 0, \\
L_F(p) &=& 2 \ln \left| \frac {p-k} {p+k} \right|.
\end{eqnarray}
The thermal correction to the bare mass is then
\begin{eqnarray}
\delta &=& g^2 C_F \int_0^{\infty} \frac {dp \, p} {(2\pi)^2} \left[ 4 n_B(p)
+ \left( 4 + \frac {2m^2} {pk} \ln \left| \frac {p-k} {p+k} \right| \right)
\frac {pn_F(E)} {E} \right], \\
&=& g^2 C_F \left[ \left( \frac 1 6 + q(\frac m T) \right) T^2
+ r (\frac m T, \frac k T) m^2 \right], \\
q (\frac m T) &=& \frac {1} {\pi^2 T^2} \int_m^{\infty} dE \, p \, n_F(E), \\
r (\frac m T, \frac k T) &=& \frac {1} {2\pi^2 \, k} \int_m^{\infty} dE \,
\ln \left| \frac {p-k} {p+k} \right| \, n_F(E).
\end{eqnarray}
It is clear from inspection that $q > 0$ and $r < 0$.


For regulation of production rates in the early stages of ultrarelativistic
nuclear collisions, the most interesting case is $\mu < 0$.  The case
$\mu > 0$ is discussed thoroughly in Refs.~[\ref{rt}] and [\ref{rbo}].  For
simplicity, I calculate the average value of $r$,
\begin{equation}
\overline{r} (\frac m T) = \frac
{\displaystyle \int \frac {d^3k} {\omega} \, r (\frac m T, \frac k T) \,
n_F(\omega)}
{\displaystyle \int \frac {d^3k} {\omega} \, n_F(\omega)},
\end{equation}
where $\omega = (K^2+m^2)^{1/2}$.
Using $\overline{r}$ for regulation of divergences is probably the most
sensible procedure, in the absence of a full calculation with effective
thermal propagators and vertices.

The most useful case for present simulations is when the gluons are
chemically equilibrated but the quark densities are very low ($-\mu/T \gg
1$), as these are the conditions under which most quark production occurs
in an ultra-relativistic nuclear collision.  In this case,
\begin{eqnarray}
\overline{\delta} (\frac m T) &=& g^2 C_F \left[ \frac {T^2} 6
+e^{\mu/T} \left( T^2 q_{\infty} (\frac m T)
+m^2 \overline{r}_{\infty} (\frac m T) \right) \right], \\
q_{\infty} (\frac m T) &=& \frac {1} {\pi^2 T^2}
\int_m^{\infty} dE \, p \, e^{-E/T}, \\
\overline{r}_{\infty} (\frac m T) &=& \frac
{\displaystyle \int_m^{\infty} d\omega \int_m^{\infty} dE \,
\ln \left| \frac {p-k} {p+k} \right| \, e^{-(E+\omega)/T}}
{\displaystyle 2\pi^2 \int_m^{\infty} d\omega \, k \, e^{-\omega/T}}.
\end{eqnarray}
The integrals can be evaluated in closed form in the low- and high-mass
limits:
\begin{eqnarray}
q_{\infty} \left( \frac m T \right) &=&
  \left\{ \begin{array}{ll}
     \frac {1} {\pi^2} \left( 1 + \frac {m^2} {2 T^2} \ln ( \frac m T )
       + {\cal O} [\frac {m^2} {T^2}] \right) \quad & m/T \ll 1, \\[2pt]
     \left( \frac {m} {2\pi^3T} \right)^{1/2} e^{-m/T} \left( 1
       + \frac {3T} {8m}
       + {\cal O} [\frac {T^2} {m^2}] \right) \quad & m/T \gg 1;
  \end{array} \right. \label{eqi} \\[3pt]
\overline{r}_{\infty} \left( \frac m T \right) &=&
  \left\{ \begin{array}{ll}
     \frac {-1} {2\pi^2} \left( 1 -\frac {m^2} {T^2}
       \ln^2 ( \frac m T ) + {\cal O} [\frac {m^2} {T^2}
       \ln ( \frac m T ) ] \right) \quad & m/T \ll 1, \\[2pt]
     -\left( \frac {T} {8 \pi^3 m} \right)^{1/2} e^{-m/T}
       \left( 1 -\frac {19T} {8m}
       +{\cal O} [\frac {T^2} {m^2}] \right) \quad & m/T \gg 1.
  \end{array} \right. \label{eri}
\end{eqnarray}
They can also be easily evaluated numerically to high accuracy.  I was
unable to find accurate interpolation formulas, probably because of the
non-analytic behavior of $q_{\infty}$ and $r_{\infty}$ at $m=0$.

\bigskip

I thank T. Altherr for useful discussions.  This material is based upon
work supported by the North Atlantic Treaty Organization under a Grant
awarded in 1991.

\vfill \eject


\ulsect{References}

\begin{list}{\arabic{enumi}.\hfill}{\setlength{\topsep}{0pt}
\setlength{\partopsep}{0pt} \setlength{\itemsep}{0pt}
\setlength{\parsep}{0pt} \setlength{\leftmargin}{\labelwidth}
\setlength{\rightmargin}{0pt} \setlength{\listparindent}{0pt}
\setlength{\itemindent}{0pt} \setlength{\labelsep}{0pt}
\usecounter{enumi}}

\item T. Bir\'o, E. van Doorn, B. M\"uller, M. Thoma and X. Wang,
Phys.\ Rev.\ C {\bf 48}, 1275 (1993). \label{rmueller}

\item K. Geiger, Phys.\ Rev. D {\bf 46}, 4965 (1992). \label{rg}

\item E. Shuryak, Phys. Rev. Lett. {\bf 68}, 3270 (1992). \label{rs}

\item T. Matsui, B. Svetitsky and L. McLerran, Phys.\ Rev.\ D {\bf 34},
783 (1986). \label{rMMS}

\item J. Rafelski and B. M\"uller, Phys.\ Rev.\ Lett.\
{\bf 48}, 1066 (1982). \label{rstrange}

\item A. Shor, Phys. Lett. {\bf B215}, 375 (1988). \label{rcharm}

\item V. Klimov, Sov.\ J. Nucl.\ Phys.\ {\bf 33}, 934 (1981);
      H.A. Weldon, Phys.\ Rev.\ {\bf D26}, 1394 (1982).  \label{rKW}

\item H.A. Weldon, Phys.\ Rev.\ D {\bf 26}, 2789 (1982). \label{rW}

\item E. Braaten and R. Pisarski, Nucl.\ Phys.\ {\bf B337}, 569 (1990);
      Nucl.\ Phys.\ {\bf B339}, 310 (1990).\label{rBP}

\item E. Levinson and D. Boal, Phys.\ Rev.\ D {\bf 31}, 3280 (1985).
\label{rlb}

\item T. Toimela, Nucl.\ Phys.\ {\bf B273}, 719 (1986). \label{rt}

\item E. Petitgirard, Zeit.\ Phys.\ C {\bf 54}, 673 (1992). \label{rpg}

\item G. Baym, J.P. Blaizot and B. Svetitsky, Phys.\ Rev.\ D {\bf 46},
4043 (1992). \label{rbbs}

\item E. Braaten, Astrophysical Journal {\bf 392}, 70 (1992).

\item J.P. Blaizot and J.Y. Ollitrault, Phys.\ Rev. D {\bf 48}, 1390
(1993). \label{rbo}

\end{list}

\vfill \eject

\end{document}